\documentclass[manuscript]{aastex}
\usepackage{graphicx}

\shorttitle{Rossiter-McLaughlin of 55 Cnc e}
\shortauthors{L\'opez-Morales et al.}

\begin{document}

%% LaTeX will automatically break titles if they run longer than
%% one line. However, you may use \\ to force a line break if
%% you desire.

\title{Rossiter-McLaughlin Observations of 55 Cnc e}

%% Use \author, \affil, and the \and command to format
%% author and affiliation information.
%% Note that \email has replaced the old \authoremail command
%% from AASTeX v4.0. You can use \email to mark an email address
%% anywhere in the paper, not just in the front matter.
%% As in the title, use \\ to force line breaks.

\author{Mercedes L\'opez-Morales}
\affil{Harvard-Smithsonian Center for Astrophysics, 60 Garden Street, Cambridge, MA 01238, USA}
\email{mlopez-morales@cfa.harvard.edu}

\author{Amaury H. M. J. Triaud\altaffilmark{1}}
\affil{Department of Physics, and Kavli Institute for Astrophysics and Space Research, Massachusetts Institute of Technology, Cambridge, MA 02139, USA}
%\email{triaud@mit.edu}

%\and
\author{Florian Rodler\altaffilmark{2}}
\affil{Max-Planck-Institut f\"ur Astronomie, K\"onigstuhl 17, D-69117 Heidelberg, Germany,}
\affil{Harvard-Smithsonian Center for Astrophysics, 60 Garden Street, Cambridge, MA 01238, USA}

\author{Xavier Dumusque\altaffilmark{1}}
\affil{Harvard-Smithsonian Center for Astrophysics, 60 Garden Street, Cambridge, MA 01238, USA}

\author{Lars A. Buchhave}
\affil{Harvard-Smithsonian Center for Astrophysics, 60 Garden Street, Cambridge, MA 01238, USA}
\affil{Centre for Star and Planet Formation, Natural History Museum of Denmark, University of Copenhagen, DK-1350 Copenhagen, Denmark}

\author{Avet Harutyunyan}
\affil{Fundaci\'on Galileo Galilei - INAF, Rambla Jos\'e Ana Fernandez P\'erez, 738712 Bre\~na Baja, Tenerife, Spain}

\author{Sergio Hoyer\altaffilmark{3}, Roi Alonso}
\affil{Instituto de Astrof\'isica de Canarias, E-38205 La Laguna, Tenerife, Spain; Dept. de Astrof\'isica, Universidad de La Laguna, E-38206 La Laguna,
Tenerife, Spain}

%\author{Bruno Chazelas}
%\affil{Observatoire Astronomique de l'Universit\'e de Gen\前ve, CH-1290 Sauverny, Switzerland}

\author{Micha\"el Gillon}
\affil{Institut d'Astrophysique et G\' eophysique, Universit\' e de Li\` ege, All\' ee du 6 Ao\^ ut 17, 4000 Li\` ege, Belgium}
%\affil{University of Li\前ge, All\'ee du 6 ao\^ut 17, Sart Tilman, Li\前ge 1, Belgium}

\author{Nathan A. Kaib}
\affil{Center for Interdisciplinary Exploration and Research in Astrophysics (CIERA) and Department of Physics and Astronomy, Northwestern University, 2131 Tech Drive, Evanston, IL 60208, USA}

\author{David. W. Latham}
\affil{Harvard-Smithsonian Center for Astrophysics, 60 Garden Street, Cambridge, MA 01238, USA}

\author{Christophe Lovis, Francesco Pepe}
%\affil{Observatoire de l悰niversit\'e de Gen\前ve, 51 chemin des Maillettes, CH-1290 Sauverny, Switzerland}
\affil{Observatoire Astronomique de l'Universit\' e de Gen\` eve, Chemin des Maillettes 51, Sauverny, CH-1290, Switzerland}

\author{Didier Queloz}
\affil{Cavendish Laboratory, J J Thomson Avenue, Cambridge CB3 0HE, UK; Observatoire Astronomique de l'Universit\' e de Gen\` eve, Chemin des Maillettes 51, Sauverny, CH-1290, Switzerland}

\author{Sean N. Raymond}
\affil{Universit\'e de Bordeaux, LAB, UMR 5804, BP F-33270 Floirac, France; CNRS, LAB, UMR 5804, F-33270 Floirac, France}

\author{Damien S\'egransan}
\affil{Observatoire Astronomique de l'Universit\' e de Gen\` eve, Chemin des Maillettes 51, Sauverny, CH-1290, Switzerland}

\author{Ingo P. Waldmann} 
\affil{Department of Physics and Astronomy, University College London, Gower Street, WC1E6BT, UK}

\author{St\'ephane Udry}
\affil{Observatoire Astronomique de l'Universit\' e de Gen\` eve, Chemin des Maillettes 51, Sauverny, CH-1290, Switzerland}

%\affil{Space Telescope Science Institute, Baltimore, MD 21218}

%% Notice that each of these authors has alternate affiliations, which
%% are identified by the \altaffilmark after each name.  Specify alternate
%% affiliation information with \altaffiltext, with one command per each
%% affiliation.

\altaffiltext{1}{Swiss National Science Foundation Fellow}
\altaffiltext{2}{Alexander-von-Humboldt Postdoctoral Fellow}
\altaffiltext{3}{Severo Ochoa Fellow}
%\altaffiltext{4}{Visiting Programmer, Space Telescope Science Institute}
%\altaffiltext{5}{Patron, Alonso's Bar and Grill}

%% Mark off your abstract in the ``abstract'' environment. In the manuscript
%% style, abstract will output a Received/Accepted line after the
%% title and affiliation information. No date will appear since the author
%% does not have this information. The dates will be filled in by the
%% editorial office after submission.

\begin{abstract}

We present Rossiter-McLaughlin observations of the transiting super-Earth 55 Cnc e collected during six transit events between January 2012 and November 2013 with HARPS and HARPS-N. We detect no radial-velocity signal above 35 cm s$^{-1}$ ($3\sigma$) and confine the stellar $v\,\sin\,i_\star$ to 0.2 $\pm$ 0.5 km s$^{-1}$. The star appears to be a very slow rotator, producing a very low amplitude  Rossiter-McLaughlin effect. Given such a low amplitude, the Rossiter-McLaughlin effect of 55 Cnc e is undetected in our data, and any spin--orbit angle of the system remains possible. We also performed Doppler tomography and reach a similar conclusion. Our results offer a glimpse of the capacity of future instrumentation to study low amplitude Rossiter-McLaughlin effects produced by super-Earths.

\end{abstract}

%% Keywords should appear after the \end{abstract} command. The uncommented
%% example has been keyed in ApJ style. See the instructions to authors
%% for the journal to which you are submitting your paper to determine
%% what keyword punctuation is appropriate.


%% From the front matter, we move on to the body of the paper.
%% In the first two sections, notice the use of the natbib \citep
%% and \citet commands to identify citations.  The citations are
%% tied to the reference list via symbolic KEYs. The KEY corresponds
%% to the KEY in the \bibitem in the reference list below. We have
%% chosen the first three characters of the first author's name plus
%% the last two numeral of the year of publication as our KEY for
%% each reference.

%% Authors who wish to have the most important objects in their paper
%% linked in the electronic edition to a data center may do so by tagging
%% their objects with \objectname{} or \object{}.  Each macro takes the
%% object name as its required argument. The optional, square-bracket 
%% argument should be used in cases where the data center identification
%% differs from what is to be printed in the paper.  The text appearing 
%% in curly braces is what will appear in print in the published paper. 
%% If the object name is recognized by the data centers, it will be linked
%% in the electronic edition to the object data available at the data centers  
%%
%% Note that for sources with brackets in their names, e.g. [WEG2004] 14h-090,
%% the brackets must be escaped with backslashes when used in the first
%% square-bracket argument, for instance, \object[\[WEG2004\] 14h-090]{90}).
%%  Otherwise, LaTeX will issue an error. 

\section{Introduction}

55 Cnc is a 0.9 $M_{\sun}$ star \citep{vonBraun2011} harbouring five known planets with masses between 0.025 $M_{\rm Jup}$ and 3.8 $M_{\rm Jup}$, and orbital periods between $\sim$ 0.74 days and  $\sim$ 4872 days \citep{Nelson2014}. The star also has a $\sim$ 0.27 $M_{\sun}$ binary companion at a projected orbital separation of 1065 AU \citep{Mugrauer2006}.

The innermost planet in the system, 55 Cnc e ($M_{\rm p}$ = 8.3 $\pm$ 0.4 $M_{\rm Earth}$; $R_{\rm p}$ = 1.94 $\pm$ 0.08 $R_{\rm Earth}$), was found to transit by \citet{Winn2011} and \citet{Demory2011}, after \citet{Dawson2010} provided a revised period of 0.74 days, shorter than the originally reported 2.8 days period by \citet{McArthur2004}. The presence of transits makes 55 Cnc e an invaluable potential target for future studies of atmospheric properties of super-Earths, and spin--orbit angle studies via the Rossiter-McLaughlin effect \citep{Rossiter1924, McLaughlin1924, Queloz2000, Gaudi2007}. In the case of spin-orbit angle studies, a result for 55 Cnc e would add to the very few measurements reported so far for Neptune and super-Earth mass planets, e.g. the detection of oblique orbits for HAT-P-11b  \citep{Winn2010,Hirano2011}, and for the multiple super-Earth systems Kepler-50 and Kepler-65 -- these last two via asteroseismology \citep{Chaplin2013}. There is also the non-detection of the Rossiter-McLaughlin effect of GJ 436 by \citet{Albrecht2012}, which indicates that the star is a very slow rotator with $v\,\sin\,i_\star < 0.4$ km s$^{-1}$.

%\citet{Kaib2011} produced numerical simulations on the 55 Cnc system to demonstrate that perturbations from the binary companion would have an impact on the spin--orbit alignment of the system and concluded that the whole planetary orbital plane is most likely misaligned, with a projected spin--orbit angle of $\sim$ 50$^{\circ}$ (assuming all the planets are in a near-coplanar, rigid body configuration). They also found that the system has a $\sim$ 30$\%$ chance of being in a retrograde configuration.

\citet{Kaib2011} showed that the distant binary companion to 55 Cnc causes the planetary system to precess as a rigid body~\citep[see also][]{innanen97,batygin11,boue14}. The planets' orbits nonetheless remain confined to a common plane such that a measurement of planet e's spin-orbit angle should be representative of the entire system. Given the unknown orientation of the wide binary orbit, \citet{Kaib2011} calculated that the plane of the planets is most likely tilted with respect to the stellar equator.  The plane could be tilted by virtually any angle, with a most probable projected spin-orbit angle of $\sim 50^\circ$ %(assuming an isotropic probability distribution for the binary orbit).  
There is also a $\sim 30\%$ chance of a retrograde configuration.  

\citet{Valenti2005} estimated a $v\,\sin\,i_\star$  for 55 Cnc of $ 2.4 \pm 0.5$ km s$^{-1}$, which, combined with the depth of the observed transit, was expected to yield a Rossiter-McLaughlin effect with a semi-amplitude of $70 \pm 15$ cm s$^{-1}$. Such an amplitude should be detectable with stabilized, high resolution spectrographs such as HARPS \citep{Molaro2013}.

Following those results we set out to detect the Rossiter-McLaughlin effect of 55 Cnc e and test the spin--orbit misalignment prediction of the system by \citet{Kaib2011}.

\section{Observations}

Shortly after the detection of 55 Cnc e's transit was announced, we requested four spectroscopic time series on HARPS  (Prog.~ID 288.C-5010; PI Triaud), as Director Discretionary Time.  HARPS is installed on the 3.6-m telescope at the European Southern Observatory on La Silla, Chile \citep{Mayor2003}. The position of 55 Cnc in the sky --- ${\rm RA(J2000)}\,$= 08:52:35.81, ${\rm Dec(J2000)}\,$= +28:10:50.95 --- is low as seen from La Silla. The target remains at a zenith distance of $z<2$ for only $\sim 2.5$ hours per night, with a transit duration of about 1.5 hours having to fit within this tight window. This constraint on the airmass, essential to obtain precise RVs, is set by the instrumental atmospheric dispersion corrector. We used the ephemeris by \citet{Gillon2012}, then at an advanced stage of preparation, to schedule our observations. In total we gathered 179 spectra on the nights starting on 2012-01-27, 2012-02-13, 2012-02-27 and 2012-03-15 UT. 

The target is better suited for observations from the north and, therefore, in 2013 we acquired additional radial velocity time series with the newly installed HARPS-N instrument on the 3.57-m Telescopio Nazionale Galileo (TNG) at the Observatorio del Roque de los Muchachos in La Palma, Spain \citep{Cosentino2012}. HARPS-N is an updated version of HARPS, and is able to reach the same overall RV precision, if not better \citep[see e.g.~][]{Pepe2013,Dumusque2014}. The HARPS-N time was awarded via the Spanish TAC (Prog.~ID 34-TNG4/13B; PI Rodler) and  observations were collected on the nights starting on 2013-11-14 and 2013-11-28 UT. We gathered a total of 113 spectra in those two nights. A third night awarded to this program was lost to weather. Some more details about the observations are provided in Table~\ref{tab:data}.

\section{Radial-Velocity extraction}

The spectra were originally reduced with version 3.7 of the HARPS and HARPS-N Data Reduction Software (DRS), which includes color systematic corrections \citep{Cosentino2014}. Radial velocities were computed using a numerical weighted mask following the methodology outlined by \citet{Baranne1996}. Such a procedure has been shown to yield remarkable precision and accuracy \citep[e.g.][]{Mayor:2009rw,Molaro2013,Pepe2013,Dumusque2014}.

However, there is still room for improvement. A detailed analysis of the RV of individual spectral lines in HARPS spectra has revealed that some lines can show variations $> 100\,{\rm m}\,{\rm s^{-1}}$ as a function of time. These variations cannot be explained by stellar noise in a star like 55 Cnc. Indeed, stellar oscillations, granulation phenomena and stellar activity are expected to be of the order of a few ${\rm m}\,{\rm s^{-1}}$ on a inactive G dwarf, as reported by \citet[][]{Dumusque2011b}. After an in-depth study of the behavior of these lines, we identify three sources of velocity errors: the stitching of the CCD, some faint telluric lines, and fringing on the detector can all introduce significant velocity shifts of certain stellar lines that happen to fall near the stationary features, changing as the barycentric velocity for different observations scans the stellar lines across the artifacts.

The spectral lines affected by these variations were identified by calculating the periodogram of the RV of each spectral line used in the stellar template. All the spectral lines that were exhibiting a signal more  significant than 10\% in false alarm probability were flagged as \emph{bad} lines and removed from the stellar template. The result is a modified mask, which is used instead of the standard K5 mask, to derive the RV of each observed spectrum by cross-correlation.

%The presence of these lines showing significant variations in the stellar template used to derive the RV by cross-correlation with the stellar spectra will induce some noise in the RV measured. 

 After performing a new cross-correlation of the data using this cleaned stellar template, the rms of the combined RV curve was reduced from $1.17\,{\rm m}\,{\rm s^{-1}}$ to $1.04\,{\rm m}\,{\rm s^{-1}}$. In the rest of our analysis we used the set of RV values from this revised reduction (a publication with this analysis is in preparation). Those radial velocities are presented in Table~\ref{tab:data} and in Figure~\ref{fig:data}.

%The spectra from each instrument were reduced with their respective Data Reduction Softwares (DRS), which has been shown to yield remarkable precision and accuracy
%\citep[e.g. {\bf Example paper for HARPS?},][]{Pepe2013,Dumusque2014}. The radial velocities were computed using a numerical weighted mask following the methodology outlined by \citet{Baranne1996}. Instead of using the standard K5 mask, we employed a modified K5 template to mitigate the influence of stellar activity, CCD stitching and CCD fringing on the radial velocities (Dumusque in prep). We compared the standard reduction with ours and only see a slight reduction of the scatter. 
%The main effect was to reduce the average error bar from $\sim95$ to $\sim75$ cm s$^{-1}$ on the HARPS sequences. 
%Our radial velocities are presented in Table~\ref{tab:data} and in Figure~\ref{fig:data}.

\section{Estimates of line broadening due to stellar rotation}\label{sec:vsini}

The large number of high signal-to-noise ratio (SNR) high-resolution spectra gathered by HARPS-N allowed us to re-determine the stellar parameters of 55 Cnc using the Stellar Parameter Classification pipeline \citep[SPC;][]{Buchhave2014}. We analyzed the 55 spectra observed on the night starting on 2013-11-14 with exposure times of 240 seconds and a resolution of R = 115,000 resulting in an average SNR per resolution element of 362 in the MgB region. We also analyzed three spectra, with SNR per resolution element of 143 and resolution R = 48,000, obtained between 2013-04-22 and 2014-03-08 with the fiber-fed Tillinghast Reflector Echelle Spectrograph \citep[TRES; ][]{Szentgyorgyi2007} on the 1.5-m Tillinghast Reflector at the Fred Lawrence Whipple Observatory on Mt. Hopkins, Arizona. The weighted average of each spectroscopic analysis yielded two sets of values from the two instruments, which are in very close agreement. The final stellar parameters are the average of these two sets, yielding $T_{\rm eff}$=5358 $\pm$ 50 K, ${\rm log(g)}$=4.44 $\pm$ 0.10, and ${\rm [m/H]}$=0.34 $\pm$ 0.08. Those values agree with the values published in the literature. The HARPS-N spectra yielded a projected rotational velocity of $v\,\sin\,i_\star = 0.43$ km s$^{-1}$, with a upper limit of 1.43 km s$^{-1}$,
and the TRES spectra gave $v\,\sin\,i_\star = 0.85$ km s$^{-1}$, with a upper limit of 1.85 km s$^{-1}$. Both values agree with each other, but lower than the value of $2.4 \pm 0.5$ km s$^{-1}$ reported by \citet{Valenti2005}. %We note that the errorbars in the reported rotational velocities include uncertainties due to the broadening of  lines by the spectrograph.

%We note, however, that the derived rotational velocities for slowly rotating stars is uncertain due to the broadening of the spectral lines by the spectrograph.

We also computed the $\rm R'_{HK}$ activity index from our spectra and derived the age and rotation period of the star following \citet{Mamajek2008}. We arrive to an age for the star of 9.3 $\pm$ 1.1 Gyr, a rotational period of $P_{\rm rot}$ = 52 $\pm$ 5 days, and a log $\rm R'_{HK}$ = -5.07 $\pm$ 0.02 using the HARPS-N spectra. The analysis of the HARPS spectra yields similar results, i.e. age = 10.0 $\pm$ 1.2 Gyr, $P_{\rm rot}$ = 54 $\pm$ 5 days, and a log $\rm R'_{HK}$ = - 5.11 $\pm$ 0.02. Both results agree with previously published values
\citep{vonBraun2011,Dragomir2014} and imply a rotational velocity slower than 1 km $s^{-1}$, in agreement with our line broadening estimates. 
%All these numbers are consistent with the star's spin axis, $i_\star$, being close to 90$^\circ$, i.e. perpendicular to the line of sight.

\section{Analysis of the radial-velocity data}

The first step of our radial velocity analysis consisted of removing the Keplerian orbital motion signals of the five planets discovered around 55 Cnc from each of our six datasets. We used the most recent orbital solution of the system obtained by \citet{Nelson2014}. %, who performed self-consistent N-body simulations of 1418 published precise radial velocity observations and transit duration and times of 55 Cnc e. 
As recommended by these authors, we used the planetary masses and orbital parameters from their {\it Case~2}, in which they considered the errors of RV observations taken within 10 minutes from each other to be perfectly correlated. The corrected radial velocities for each dataset are given in Table~\ref{tab:data}. Later tests using the \citet{Nelson2014} system parameters for their {\it Case 1}, and also the planetary masses and orbital parameters of the system derived by \citet{Dawson2010} in their tables 7, 8 and 10, yield similar results.

After removing the five-planet signal, we modelled the Rossiter-McLaughlin effect for the six datasets combined using the formalism of \citet{Gimenez2006}, based on \citet{Kopal1942}, and adjusted with a Monte Carlo Markov-Chain described in \citet{Triaud2013}. All parameters determined by the photometry were controlled by priors issued from two independent datasets (see Table~\ref{tab:results}). Additional priors included the planet's period \citep{Nelson2014}, the stellar parameters \citep{vonBraun2011}, and the two $v\,\sin\,i_\star$ values estimated in Sect.~\ref{sec:vsini}. We allowed the relative mean of each time series to float in order to absorb any offset produced by slightly different epochs of stellar activity \citep[e.g.~][] {Triaud2009}. Thirteen parameters were thus used on a total of 293 data points. A noise term of 70 cm s$^{-1}$ was quadratically added to all measurement errors to reach a final reduced $\chi^2$ of $0.97\pm0.08$. All priors and important results are summarized in Table~\ref{tab:results}.

\begin{figure}
\epsscale{1.2}
%\plottwo{RVdata.eps}{RVbin.eps}
\plottwo{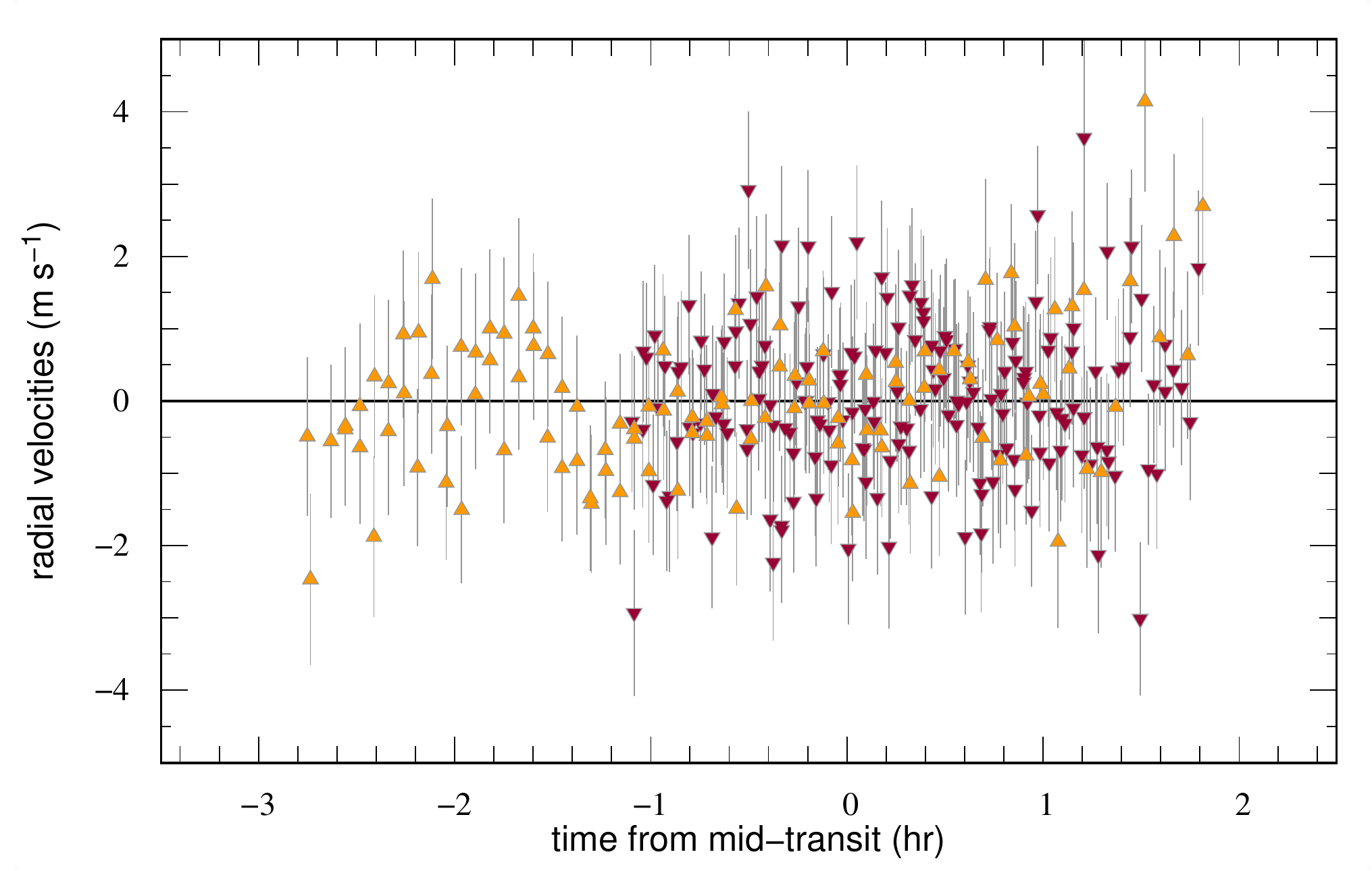}{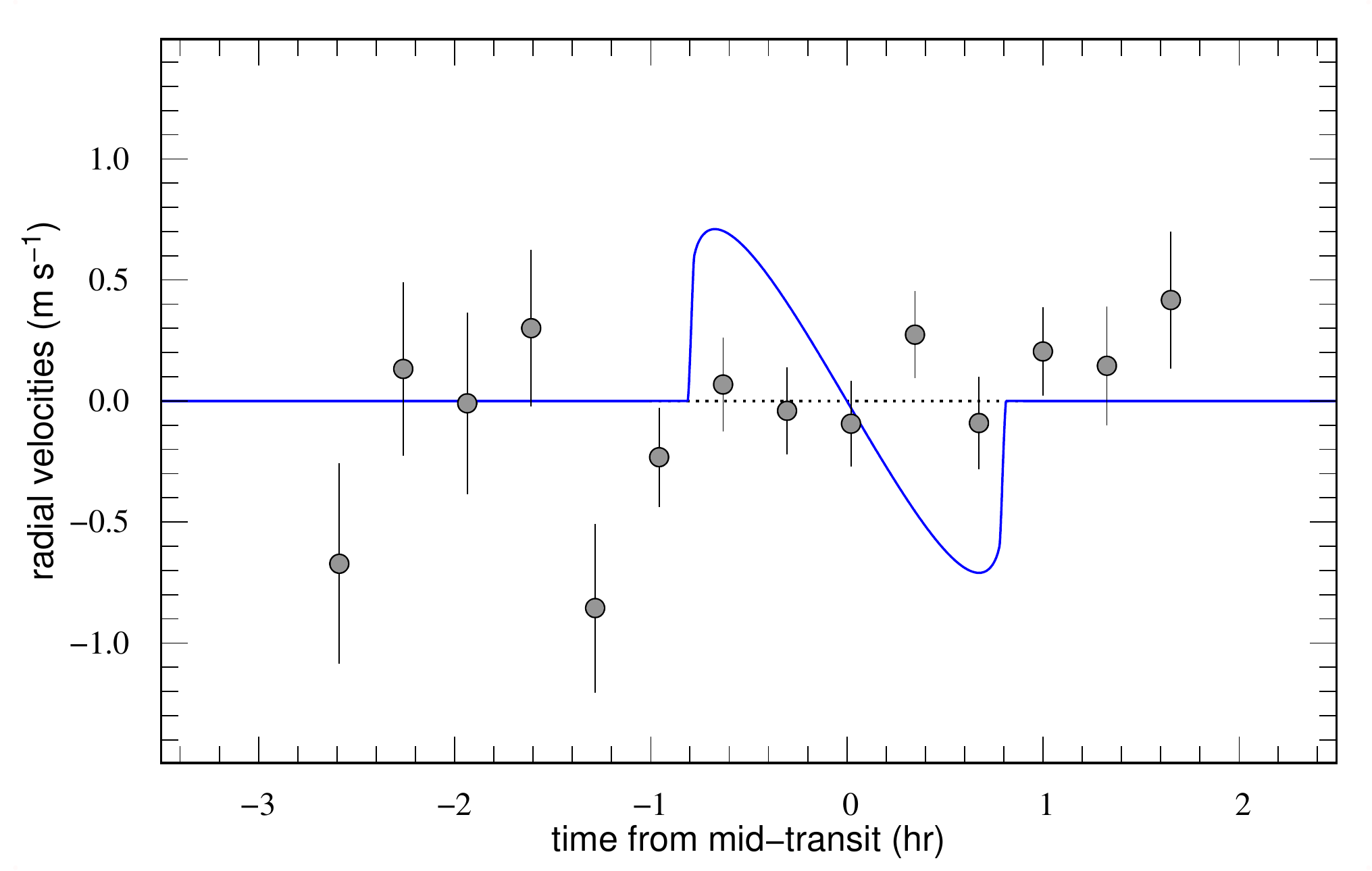}
\caption{Left: Radial-velocity data, in phase. HARPS data points are represented by inverted, red triangles, and HARPS-N by upright, orange triangles. Right: the same data, binned in 14 equidistant points , and a model of the Rossiter-McLaughlin effect, for $v\,\sin\,i_\star = 2.4 $ km s$^{-1}$ and $\beta = 0^\circ$.\label{fig:data}}
\end{figure}

\begin{figure}
\includegraphics[scale=0.75]{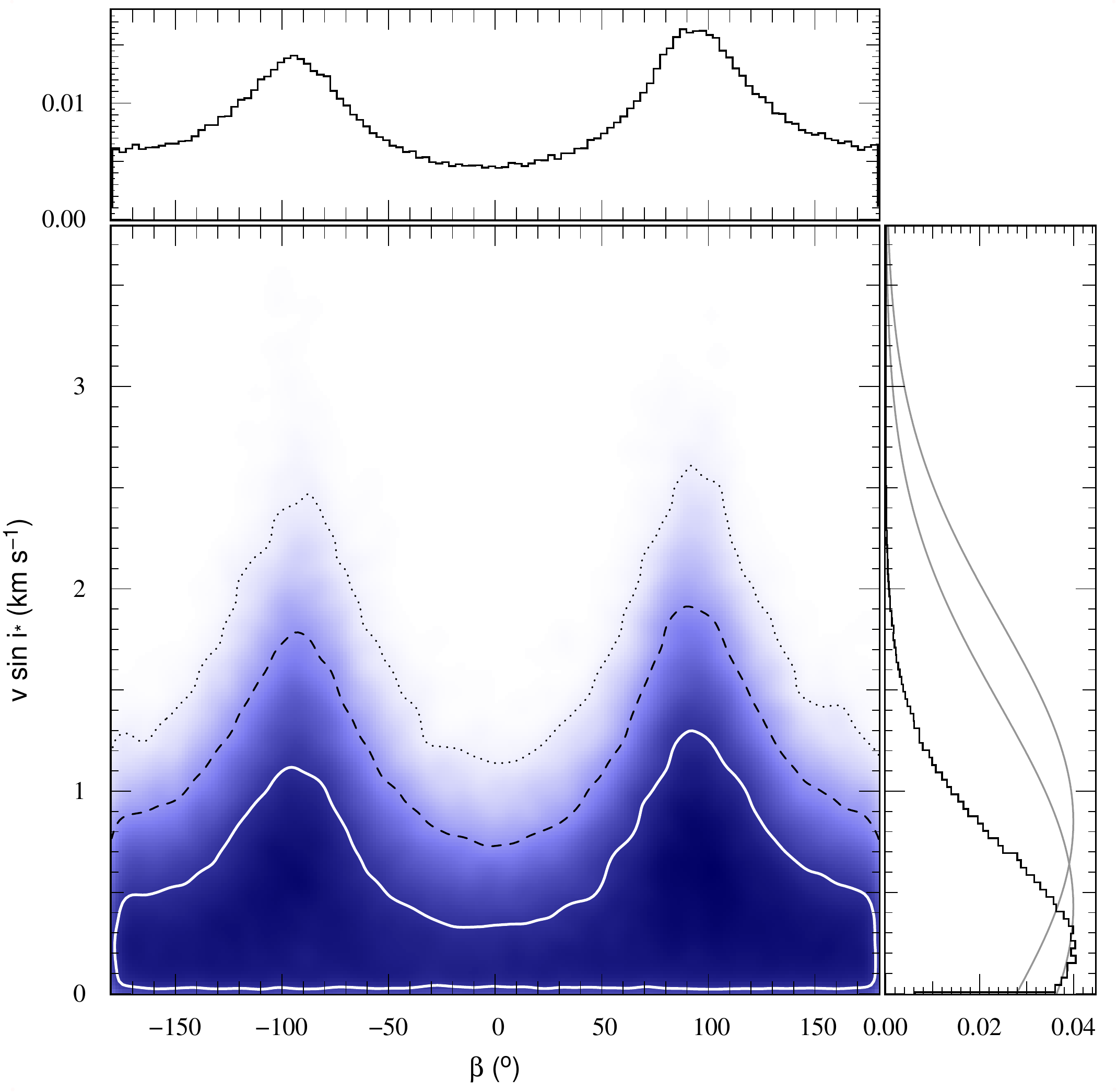}
\caption{Posterior distribution for the projection of the stellar rotational speed, $v\,\sin\,i_\star$, and the projection on the sky of the spin--orbit angle, $\beta$. 1, 2 and 3 $\sigma$ confidence contours are over plotted. Side histograms display the marginalised posteriors for each quantity. In the case of $v\,\sin\,i_\star$, the two priors we applied are drawn as grey lines.\label{fig:posterior}}
\end{figure}

We conducted a number of different chains all converging to the same conclusion: the Rossiter-McLaughlin effect is not detected (see Figure~\ref{fig:data}). The impossibility to constrain the spin--orbit angle of the system is illustrated in Figure~\ref{fig:posterior}, %, which shows the posterior distribution in ($v\,\sin\,i_\star$, $\beta$) space, where $\beta$ is the projection on the sky of the spin--orbit angle. 
which shows the typical crescent shape expected when there is degeneracy between fast rotation \& polar orbits and slow rotation \& alignment \citep{Triaud2011, Albrecht2011}. This shape approximately maps contours of Rossiter-McLaughlin effects of equal semi-amplitudes. From the $3 \sigma$ contour, we rule-out Rossiter-McLaughlin effects with semi-amplitudes larger than 35~cm~s$^{-1}$. %-- a feat in itself. 
The fact that anti-aligned orbits are as likely as aligned orbits implies the Rossiter-McLaughlin effect is not detected. 

The marginalised distribution in $v\,\sin\,i_\star$ is thinner and peaks closer to 0 than our two estimates from spectral line broadening. Therefore, from the data we estimate $v\,\sin\,i_\star = 0.2 \pm0.5$ km s$^{-1}$. This implies the star's most likely rotation period is 260 days ($>22$ days, with $3\sigma$ confidence; $> 40$ days, only considering coplanar solutions).

To test the robustness of our analyses, we explored the impact of different priors. When replacing our two $v\,\sin\,i_\star$ priors with the value estimated in \citet{Valenti2005}, two symmetrical solutions, on polar orbits, are preferred: $\beta = +90^\circ$ and $-90^\circ$ (such a Rossiter-McLaughlin effect would have a semi-amplitude of 23 cm~s$^{-1}$). A similar situation occurred for the spin--orbit angle measurement of WASP-80b \citep{Triaud2013}, which depends entirely on the value of $v\,\sin\,i_\star$. Using no priors on $v\,\sin\,i_\star$, the posterior is qualitatively similar, but both spikes are thinner and $v\,\sin\,i_\star$ tails to higher values, as would be expected. The same procedures, removing the additional 70 cm s$^{-1}$ noise added to the errorbars, produced similar results (with shorter confidence intervals).

%\begin{figure}
%\epsscale{1.2}
%\plottwo{RVdata.eps}{RVbin.eps}
%\plottwo{RVdata.pdf}{RVbin.pdf}
%\caption{Left: Radial-velocity data, in phase. HARPS data points are represented by inverted, red triangles, and HARPS-N by upright, orange triangles. Right: the same data, binned for visual convenience in 14 equidistant points , and a model of the Rossiter-McLaughlin effect, for $v\,\sin\,i_\star = 2.4 $ km s$^{-1}$ and $\beta = 0^\circ$. \label{fig:data}}
%\end{figure}

%\begin{figure}
%%\includegraphics[scale=0.75]{55cnc_angle_vsini_prior3.eps}
%\includegraphics[scale=0.75]{55cnc_angle_vsini_prior3.pdf}
%\caption{Posterior distribution for the projection of the stellar rotational speed, $v\,\sin\,i_\star$, and the projection on the sky of the spin--orbit angle, $\beta$. 1, 2 and 3 $\sigma$ confidence contours are over plotted. Side histograms display the marginalised posteriors for each quantity. In the case of $v\,\sin\,i_\star$, the two priors we applied are drawn as grey lines.\label{fig:posterior}}
%\end{figure}

\section{Doppler tomography}

Given the impossibility to detect the Rossiter-McLaughlin effect of 55 Cnc e, we tried to detect the signal of the planet using Doppler tomography \citep[e.g.][]{Albrecht:2007th, Collier-Cameron:2010lk}. Doppler tomography reveals the distortion of the stellar line profiles when the planet, during transit, blocks part of the stellar photosphere. This distortion is a tiny dip in the stellar absorption profile, scaled down in width accordingly to the planet-to-star radius ratio. Additionally, the area of that dip corresponds to the planetary-to-stellar disks area ratio. As the planet moves across the stellar disk, the dip produces a trace in the time series of line profiles, which reveals the spin-orbit alignment between star and planetary orbit.

%In the observed stellar spectrum, the fraction of the missing starlight manifests itself in a deformation of the stellar absorption lines. By comparing stellar absorption lines out of and within transit, it is possible to recover the signal caused by the planet. %{\bf [previous sentence looks like the traditional RM, not doppler imagin]} 
%The distortion of the stellar line profile leads to small radial velocity variations that can be detected using cross-correlation techniques. By tracing those small radial velocity variations we can trace the transiting planet.

\begin{figure}
\includegraphics[scale=0.75, angle=270]{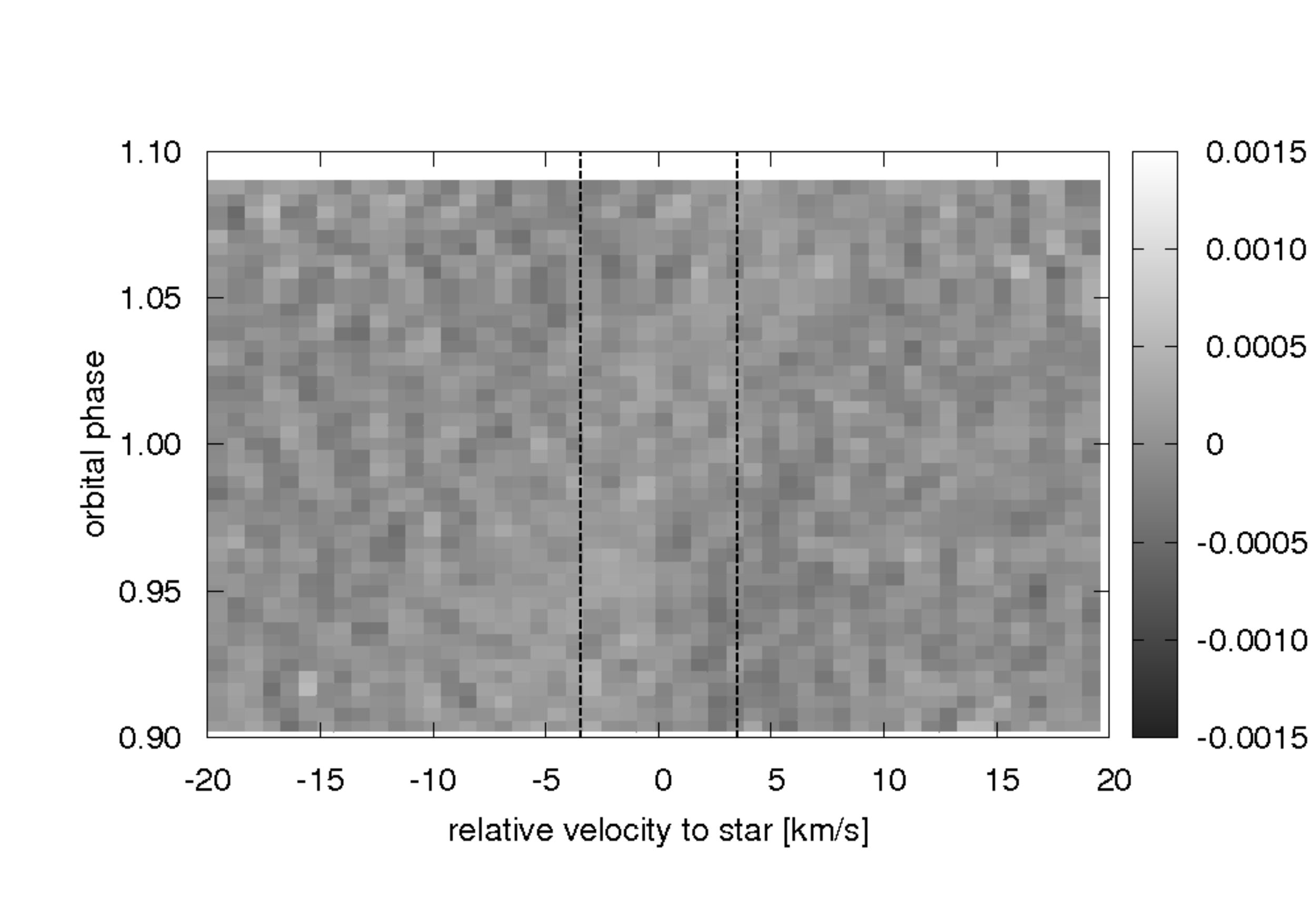}
\caption{Residuals of all line profiles of 55 Cnc taken during the six transits as a function of velocity and orbital phase of the planet. The two vertical dashed lines depict the area of the stellar line profile (FWHM).  The units of the grey scale are fractional deviation from the average out-of-transit line profile. We are unable to detect the planetary signature in the line profiles.\label{fig:dimaging}}
\end{figure}

For this analysis we summed up all the thousands of stellar absorption lines in each spectrum into one high S/N mean line profile. For this step, we employed a least-squares deconvolution \citep[LSD; e.g.][]{CollierCameron2002} of the observed spectrum and theoretical lists of the stellar absorption lines from the Vienna Atomic Line Database \citep[VALD,][]{Kupka1999} for a star with  $T_{\rm eft}=5200$~K and $\log g = 4.5$. 
The resulting line profiles were scaled so that their height was one, and were interpolated onto a velocity grid of 0.79~km~s$^{-1}$ increments, corresponding to the velocity range of one spectral pixel of HARPS-N at 550~nm. We then corrected the stellar line profiles for the radial velocity of the host star and the barycentric velocity of the Earth. For each of the six runs, we summed up all the mean line profiles collected before and after the transit and subtracted the resulting profile from the in-transit ones. We then sorted the co-aligned line profiles of all runs by orbital phase and combined the line profiles into one dataset. Figure~\ref{fig:dimaging} shows the residuals of the line profiles and demonstrates that we are also unable to detect a trace of the transiting planet using this method. Our ability to detect the planet using Doppler tomography is in fact limited by the resolution of the spectra ($\sim$ 2.6 ~km~s$^{-1}$  per resolution element), given the very slow rotational velocity of the star. %However, had the star been rotating at 2.4 km$s^-1$ as reported by \citet{Valenti2005}, or faster, we would have clearly detected a Doppler shift of  70~cm~s$^{-1}$, as originally estimated.

\section{Conclusions}

Our data do not support a detection of the spectroscopic transit of 55 Cnc e, but they rule out with $3\sigma$ confidence any signal with a semi-amplitude larger than 35~cm~s$^{-1}$.   The non-detection %of the effect of the planet using either the Rossiter-McLaughlin anomaly or Doppler tomography 
can be explained by one of three scenarios: either 55 Cnc rotates more slowly than the average G-K main sequence star \citep[$\sim$ 2.4 km~${\rm s^{-1}}$; ][]{Valenti2005}, or the orbital plane of the planets is perpendicular to the equatorial plane of the star, or the spin axis of the star is highly inclined with respect to us.
\citet{Dragomir2014} report no variability associated to stellar spots rotation on a 42 day continuous monitoring of 55 Cnc with {\it MOST}. %No variability associated by stellar spots rotating in and out of view was detected. 
\citet{vonBraun2011} estimated the system age at  $10.2\pm2.5$ Gyr. Therefore, the star is likely a very slow rotator, which is confirmed by our chromospheric activity measurements and stellar line broadening. Our conclusion is also consistent with a photometric modulation of $42.7\pm2.5$ days reported by \citet{Fischer2008}, which they interpreted as stellar rotation. Because of the slow stellar rotation, reliably determining the projected spin--orbit angle of 55 Cnc e, or the inclination of its host's spin axis may remain out of reach. The slow stellar rotation also implies a weaker tidal coupling between 55 Cnc e and its host than what presumed by \citet{boue14}. Following \citet{Kaib2011}, that would increase the probability of a spin--orbit misalignment. 

While the detection of such a misalignment currently remains out of reach, our analysis confirms the capacity of the HARPS technology to reach a few tens of cm~s$^{-1}$. These radial-velocity time series are the most constraining yet in precision and they reveal the suitability of HARPS and HARPS-N for follow-up and confirmation of small planets  to be detected by missions like TESS. Such level of radial velocity precision will also be routinely reached by ESPRESSO on the VLT. ESPRESSO will open up the study of weak Rossiter-McLaughlin effects, produced by super-Earths transiting slow rotators such as 55 Cancri e, with the potential to see if the same diversity that has been observed in the spin--orbit angle of hot Jupiters \citep{Triaud2010, Brown2012, Albrecht2012} also exists for small planets.

%{\it Addendum:}  Just before we submitted this paper, Bourrier \& H\'ebrard (2014) published a detection of the Rossiter-McLaughlin effect of 55 Cnc e. The results of our analysis disagree with their claim.

Finally, Bourrier \& H\'ebrard (2014) recently reported a detection of the Rossiter-McLaughlin effect of 55 Cnc e, with an amplitude of $\sim$ 60~cm~s$^{-1}$. While such radial velocity variations could be attributed to stellar surface physical phenomena (e.g. granulation or faculae), we argue that they are not produced the Rossiter-McLaughlin effect, since the necessary $v\,\sin\,i_\star$ $\sim$ 3.3 km s$^{-1}$ is incompatible with the $v\,\sin\,i_\star$, $P_{\rm rot}$, age, and $\rm R'_{HK}$ activity index values we measure in Sect.~4. That $v\,\sin\,i_\star$ is also incompatible with the $\sim$ 10 Gyr stellar age derived by \citet{vonBraun2011}, and with the $P_{\rm rot} > 40$ days rotational period estimates from \citet{Fischer2008} and \citet{Dragomir2014}.

\acknowledgments

This publication was made possible through the support of a grant from the John Templeton Foundation. The opinions expressed are those of the authors and do not necessarily reflect the views of the John Templeton Foundation. The research leading to these results received funding from the European Union Seventh Framework Programme (FP7/2007-2013) under Grant Agreement 313014 (ETAEARTH). This work used the VALD database, operated at Uppsala University, the Institute of Astronomy RAS in Moscow, and the University of Vienna. We thank the anonymous referee for a very constructive review and J. Winn for useful comments. A. H.\,M.\,J.\,T. is a Swiss National Science Foundation fellow under grant number P300P2-147773. 
F.\,R. acknowledges funding from the Alexander-von-Humboldt postdoctoral fellowship program. X.\,D. thanks the Swiss National Science Foundation (SNSF) for its support through an Early Postdoc Mobility fellowship.  S.\,H. acknowledges support from the Spanish Ministry of Economy and Competitiveness under the 2011 Severo Ochoa Program MINECO SEV-2011-0187. HARPS-N is a collaboration between the Astronomical Observatory of the Geneva University, the CfA in Cambridge, the University of St. Andrews, the Queens University of Belfast, and the TNG-INAF Observatory. We thank all the researchers whose observing programs on HARPS got disrupted by our DDT program and who kindly observed for us. Finally we thank ESO's director general, Tim de Zeeuw, who granted our HARPS observing time.

{\it Facilities:} \facility{ESO:3.6m:HARPS}, \facility{TNG:HARPS-N}.

%% Appendix material should be preceded with a single \appendix command.
%% There should be a \section command for each appendix. Mark appendix
%% subsections with the same markup you use in the main body of the paper.

%% Each Appendix (indicated with \section) will be lettered A, B, C, etc.
%% The equation counter will reset when it encounters the \appendix
%% command and will number appendix equations (A1), (A2), etc.

%\appendix

%\section{Appendix material}

%% We have used macros to produce journal name abbreviations.
%% AASTeX provides a number of these for the more frequently-cited journals.
%% See the Author Guide for a list of them.

%% Note that the style of the \bibitem labels (in []) is slightly
%% different from previous examples.  The natbib system solves a host
%% of citation expression problems, but it is necessary to clearly
%% delimit the year from the author name used in the citation.
%% See the natbib documentation for more details and options.

%\bibliographystyle{apj}
%\bibliography{ref.bib}

\clearpage

%% Use the figure environment and \plotone or \plottwo to include
%% figures and captions in your electronic submission.
%% To embed the sample graphics in
%% the file, uncomment the \plotone, \plottwo, and
%% \includegraphics commands
%%
%% If you need a layout that cannot be achieved with \plotone or
%% \plottwo, you can invoke the graphicx package directly with the
%% \includegraphics command or use \plotfiddle. For more information,
%% please see the tutorial on "Using Electronic Art with AASTeX" in the
%% documentation section at the AASTeX Web site, http://aastex.aas.org/
%%
%% The examples below also include sample markup for submission of
%% supplemental electronic materials. As always, be sure to check
%% the instructions to authors for the journal you are submitting to
%% for specific submissions guidelines as they vary from
%% journal to journal.

%% This example uses \plotone to include an EPS file scaled to
%% 80% of its natural size with \epsscale. Its caption
%% has been written to indicate that additional figure parts will be
%% available in the electronic journal.

%\begin{figure}
%\epsscale{.80}
%\plotone{f1.eps}
%\caption{Derived spectra for 3C138 \citep[see][]{heiles03}. Plots for all sources are available
%in the electronic edition of {\it The Astrophysical Journal}.\label{fig1}}
%\end{figure}

\clearpage

\clearpage

\begin{deluxetable}{lrrrrrrrlr}
\tabletypesize{\scriptsize}
\tablecolumns{10}
\tablewidth{0pt}
\tablecaption{Radial velocities of all spectra of 55 Cnc observed with HARPS and HARPS-N. The columns are:
barycentric Julian date, radial velocity,  radial velocity errorbars, full width at half maximum of the cross-correlation function, span in bisector slope, as defined in \citet{Queloz2000}, signal- to-noise ratio at 6240 ${\rm \AA}$, airmass, seeing measured by the La Silla seeing monitor, exposure time, and radial velocities with the five planet Doppler displacement removed$^{a}$.\label{tab:data}}
\tablehead{
\colhead{BJD}           & \colhead{RV}      & \colhead{$\sigma_{\rm RV}$}          & \colhead{FWHM}  & \colhead{BIS$_{\rm span}$}          & \colhead{S/N}    & \colhead{airmass}  &\colhead{seeing}  & \colhead{exposure} & \colhead{RV$_{\rm cor}$}\\

\colhead{$-2\,400\,000$}           & \colhead{km s$^{-1}$}      & \colhead{km s$^{-1}$}          & \colhead{km s$^{-1}$}  & \colhead{km s$^{-1}$}          & \colhead{at $6240 {\rm \AA}$ }    & \colhead{\nodata}  & \colhead{arcsec}  & \colhead{s} & \colhead{km s$^{-1}$}
}
\startdata
%\multicolumn{1}{l}{HARPS: 2012-01-27}\\
%\\
55954.663365&27.42076&0.00090&6.22245&-0.01210&118.80&2.04&0.74&123.1&-0.00298\\ 
55954.665229&27.41805&0.00082&6.22627&-0.00865&133.50&2.03&0.81&123.1&-0.00044\\ 
55954.667405&27.41877&0.00066&6.22736&-0.00984&181.50&2.01&0.78&180.0&-0.00121\\ 
55954.669905&27.41707&0.00066&6.22624&-0.01121&178.90&2.00&0.72&180.0&0.00044\\ 
55954.672474&27.41806&0.00065&6.22813&-0.01089&182.80&1.98&0.76&180.0&-0.00061\\ 
\nodata             & \nodata   & \nodata & \nodata & \nodata & \nodata & \nodata & \nodata & \nodata & \nodata \\
\\
\enddata
\tablenotetext{a}{The full table is published in the journal's electronic edition. A portion is reproduced here to show its form and content.}
%% You can append references to a table using the \tablerefs command.
\end{deluxetable}

%\clearpage

\begin{deluxetable}{lrrrr}
\tabletypesize{\scriptsize}
\tablecolumns{5}
\tablewidth{0pt}
\tablecaption{Analysis priors and results. The quantities determined by photometry are inserted as Gaussian priors taken from three papers. The values our analysis yields are included to illustrate that our fits did not force solutions to unrealistic values.\label{tab:results}}
\tablehead{
\colhead{Quantities [units]}           & \colhead{ \citet{Winn2011}}      & \colhead{\citet{Demory2011}}          & \colhead{\citet{Nelson2014}}  & \colhead{this paper}\\

}
\startdata
Transit depth, $\Delta F$ [ppm]			& $380\pm52$				& $410\pm63$				& \nodata	& $384\pm41$	\\
Impact parameter, $b$ [R$_\star$]		& $0.00\pm0.24$			& $0.16\pm0.13$			& \nodata & $0.18\pm0.08$\\
Transit duration, $W$ [d]				& $0.658\pm0.0019$		& $0.0665\pm0.0019$		& \nodata & $0.0667\pm0.0008$\\
Mid-transit time, $T0$ [BJD-2\,450\,000]	&$ 6184.50910\pm0.00087$	& $6184.5170\pm0.0015$	& \nodata & $6184.5120\pm0.0011$\\
Period, $P$ [d]						& \nodata					& \nodata					& $0.7365478\pm0.0000014$	& $0.7365478\pm0.0000014$\\
$\sqrt{V\sin i_\star} \cos \beta$			& \nodata					& \nodata					& \nodata & $-0.06\pm0.42$	\\
$\sqrt{V\sin i_\star} \sin \beta$			& \nodata					& \nodata					& \nodata & $0.06\pm0.69$	\\
\tableline
Stellar $v \sin i_\star$ [km s$^{-1}$]		& \nodata					& \nodata					& \nodata & $ 0.18 \pm 0.48$\\
Projected spin--orbit angle $\beta$ [deg]	& \nodata					& \nodata					& \nodata & $ 0 \pm 180$\\
Stellar rotation period, $P_{\rm rot}$ [d]	& \nodata					& \nodata					& \nodata & $> 20$ ($3\sigma$)\\
\enddata
%\tablenotetext{a}{maybe add the median seeing, signal to noise and error bar?}
%% You can append references to a table using the \tablerefs command.
\end{deluxetable}

%% Tables may also be prepared as separate files. See the accompanying
%% sample file table.tex for an example of an external table file.
%% To include an external file in your main document, use the \input
%% command. Uncomment the line below to include table.tex in this
%% sample file. (Note that you will need to comment out the \documentclass,
%% \begin{document}, and \end{document} commands from table.tex if you want
%% to include it in this document.)

%% \input{table}

%% The following command ends your manuscript. LaTeX will ignore any text
%% that appears after it.

\end{document}